\documentclass[a4paper,11pt]{article}
\pdfoutput=1 % if your are submitting a pdflatex (i.e. if you have
\usepackage{jheppub} % for details on the use of the package, please
\usepackage[T1]{fontenc} % if needed

\usepackage{epsfig}
\usepackage[normalem]{ulem}
\usepackage{tikz}
\usetikzlibrary{arrows}
\usetikzlibrary{decorations.markings}
\usepackage{multirow}
\usepackage{makecell}
\usepackage{dsfont}

\newcommand{\beq}{\begin{equation}}
\newcommand{\eeq}{\end{equation}}
\newcommand{\bal}{\begin{aligned}}
\newcommand{\eal}{\end{aligned}}
\newcommand{\rmd}{\mathrm{d}}

\title{\boldmath From Rotating to Charged Black Holes and Back Again}

\author{Lars Aalsma,}
\author{Gary Shiu}

\affiliation{Department of Physics, University of Wisconsin-Madison, 1150 University Avenue, Madison, Wisconsin 53706, USA}

\emailAdd{laalsma@wisc.edu}
\emailAdd{shiu@physics.wisc.edu}

\abstract{The mild form of the Weak Gravity Conjecture (WGC) requires higher derivative corrections to extremal charged black holes to increase their charge-to-mass ratio. This allows decay via emission of a smaller extremal black hole. In this paper, we investigate if similar constraints hold for extremal rotating black holes. We do so by considering the leading higher derivative corrections to the four-dimensional Kerr black hole and five-dimensional Myers-Perry black hole. We use a known mapping of these rotating solutions to a four-dimensional non-rotating dyonic Kaluza-Klein black hole and impose the WGC on this charged solution. Going back again to the rotating solutions, this fixes the sign of the corrections to the rotating extremality bounds. The sign of the corrections is non-universal, depending on the black hole under consideration. We argue that this is not at odds with black hole decay, because of the presence of a superradiant instability that persists in the extremal limit. When this instability is present, the WGC is implied for the four-dimensional charged black hole.
}

\begin{document}

\maketitle
\flushbottom

\section{Introduction}
\label{sec:intro}
According to the Weak Gravity Conjecture (WGC), black holes should be able to decay even in the extremal limit where their temperature goes to zero and decay via Hawking radiation shuts off \cite{Arkani-Hamed:2006emk}. Although this motivation is somewhat heuristic and difficult to substantiate, it provides a useful intuition that is helpful in investigating various swampland conjectures. Indeed, the instability of gravitational solutions seems to be a key property of quantum gravity in the absence of stabilizing symmetries such as supersymmetry \cite{Ooguri:2016pdq,Freivogel:2016qwc}. Of course, most rigorous evidence in favor of the WGC comes from string theory compactifications. In fact, string theory considerations have motivated stronger versions of the WGC \cite{Heidenreich:2016aqi,Montero:2016tif,Andriolo:2018lvp} which require infinitely many superextremal states. By now, numerous tests of the WGC, including its strong forms, have been carried out in different string compactifications \cite{Lee:2018urn,Lee:2018spm,Lee:2019tst,Aalsma:2019ryi,Gendler:2020dfp,Klaewer:2020lfg}. No single violation has been found to date substantiating the heuristic black hole decay argument.

Various works have therefore applied this logic to different spacetimes, such as charged black holes in asymptotically Anti-de Sitter space \cite{Harlow:2015lma,Montero:2018fns} and de Sitter space \cite{Montero:2019ekk,Montero:2021otb}, which gives new constraints on the particle spectrum of a theory for it to belong to the landscape. This idea has also been applied to bound the sign of higher derivative corrections to extremal black holes \cite{Kats:2006xp,Cheung:2018cwt,Hamada:2018dde,Cremonini:2019wdk,Jones:2019nev,Cano:2019ycn,Cano:2019oma,Cremonini:2020smy,Cano:2021nzo}. Satisfying this so-called mild version of the WGC requires the sign of the Wilson coefficients to increase the charge-to-mass ratio of extremal black holes. Vice versa, positivity bounds on Wilson coefficients have been used to give derivations of the WGC \cite{Hamada:2018dde,Bellazzini:2019xts,Arkani-Hamed:2021ajd}, and such positivity arguments have been extended to different classes of black holes \cite{Loges:2019jzs,Loges:2020trf}.

It should be noted that the scattering positivity bounds are weakened by gravity \cite{Hamada:2018dde, Alberte:2020jsk,Tokuda:2020mlf}, and so their applicability to proving the WGC holds under the assumption that the gravitational modification is small \cite{Hamada:2018dde, Henriksson:2022oeu}. Higher derivative corrections to purely rotating solutions have also been considered \cite{Cano:2019ore,Reall:2019sah,Aalsma:2021qga}.\footnote{Higher derivative corrections have also been found to decrease the Euclidean action of axionic wormholes with a fixed charge \cite{Andriolo:2020lul,Andriolo:2022rxc}. Moreover, pure axionic wormholes were recently shown to be perturbatively stable \cite{Loges:2022nuw}. These results combined seems to suggest that the axionic WGC is a statement about wormhole fragmentation dominating the path integral}

At first sight, requiring black hole decay does not seem to yield any constraints on rotating black holes. Extremal rotating black holes can lose their angular momentum via superradiance. This instability allows these black holes to lose their angular momentum without a constraint on the angular momentum-to-mass ratio that bears resemblance to the WGC with charge exchanged for angular momentum. Moreover, for pure gravity in six and higher spacetime dimensions, black holes with a fixed mass may have arbitrarily large angular momentum \cite{Myers:1986un}. On the other hand, assuming a weakly coupled UV completion, coupling a tower of higher-spin states to gravity does seem to impose a WGC-like constraint on the light particle spectrum\footnote{More precisely, the WGC-like constraint applies to higher spin states with mass $m_J \ll \Lambda_{\rm QFT}$ where $\Lambda_{\rm QFT}$ is the energy scale where QFT breaks down.} for the theory to be causal \cite{Kaplan:2020ldi}. Furthermore, it was shown that for BTZ black holes perturbed by a relevant deformation, a spinning WGC follows from the holographic c-theorem \cite{Aalsma:2020duv}. Overall, these arguments suggest that the status of a rotating WGC is unclear.

In this paper, we will further study the rotating WGC by focussing on a particular class of rotating black hole solutions in four and five dimensions, that can be mapped to charged and non-rotating four-dimensional Kaluza-Klein black holes. This mapping has the advantage that, without rotation, the WGC can be unambiguously imposed on the four-dimensional Kaluza-Klein black hole. By including the leading higher derivative corrections to these black holes and imposing the WGC, we obtain new positivity bounds that need to be satisfied by extremal four-dimensional Kerr and five-dimensional Myers-Perry black holes with two equal rotations. This can be viewed as the total landscaping principle \cite{Aalsma:2020duv} on steroids: we fix the sign of the Wilson coefficients by imposing the WGC on one black hole solution and see how this constraint propagates under an intricate chain of dualities. The logic we apply here is similar to that of \cite{Cremonini:2020smy}. In that paper, black holes with a NUT charge were constrained by reducing to one dimension lower where this charge becomes magnetic. What is even more interesting here, as we shall see, is that mapping rotation to charge allows us to fix the sign of a Wilson coefficient whose positivity bound from scattering is currently unknown. Similar relations between charged and rotating black holes have been exploited in \cite{Aalsma:2020duv}.

Our main finding is that the sign of the corrections to the extremality bound of rotating black holes do not seem to fall into a pattern consistent with a rotating WGC bound. We show in detail that this is consistent with the black hole decay argument, because, as anticipated, these black holes have a superradiant instability. However, satisfying the charged WGC still constrains the sign of the Wilson coefficients since this instability can only occur when the charged WGC is satisfied.

The rest of this paper is organized as follows. In section \ref{sec:RotChargeMap} we show how the Kerr and five-dimensional Myers-Perry black hole can be mapped to a non-rotating dyonic Kaluza-Klein black hole. Then, in section \ref{sec:HDcorrections} we compute the leading higher derivative corrections to the extremality bound of extremal Kerr, Myers-Perry and non-rotating dyonic Kaluza-Klein black holes. We discuss the relation between superradiance and the WGC in section \ref{sec:superradiance} and conclude in section \ref{sec:conclusion}.

\section{Mapping Rotation to Charge} \label{sec:RotChargeMap}

\subsection{Kaluza-Klein Black Hole}
To map extremal rotating Kerr and five-dimensional Myers-Perry black holes to pure charge solutions, we start with a rotating five-dimensional black hole constructed in \cite{Larsen:1999pp} that is a solution to the following five-dimensional action.
\beq
I = \frac1{16\pi G_5}\int\rmd^5x\sqrt{-g_5}R_5 ~.
\eeq
The line element of interest is (following the presentation of \cite{Emparan:2007en}) given by
\beq \label{eq:5DKKBH}
\rmd s^2 = \frac{H_q}{H_p}(\rmd y + {\bf A})^2 - \frac{\Delta_\theta}{H_q}\left(\rmd t-{\bf B}\right)^2 + H_p\left(\frac{\rmd r^2}{\Delta}+\rmd\theta^2 +\frac{\Delta}{\Delta_\theta}\sin^2\theta\rmd\phi^2\right) ~.
\eeq
The explicit expressions appearing in this metric are quite lengthy and given by
\beq
\bal
H_p  &=  r^{2}+\alpha^{2}\cos^{2}{\theta}+r(p-2m)+\frac{p}{p+q}\frac{(p-2m)(q-2m)}{2} \\
&+\frac{p}{2m(p+q)}\sqrt{(q^{2}-4m^{2})(p^{2}-4m^{2})}\,\alpha\cos{\theta} ~, \\
H_q  &=  r^{2}+\alpha^{2}\cos^{2}{\theta}+r(q-2m)+\frac{q}{p+q}\frac{(p-2m)(q-2m)}{2} \\
& -\frac{q}{2m(p+q)}\sqrt{(q^{2}-4m^{2})(p^{2}-4m^{2})}\,\alpha\cos{\theta} ~,
\eal
\eeq
\beq
\bal
\Delta& =  r^{2}+\alpha^2-2mr ~, \\
\Delta_{\theta} & =  r^{2}+\alpha^{2}\cos^{2}{\theta}-2mr ~, 
\eal
\eeq
\beq
\bal
\mathbf{A}  &=  -\left[\frac{Q}{4\pi}(r+\frac{p-2m}{2})-\sqrt{\frac{q^{3}(p^{2}-4m^{2})}{4m^{2}(p+q)}}\;\alpha\cos{\theta}\right]H_{q}^{-1}\rmd t- \Bigg[\frac{P}{4\pi}(H_{q}+\alpha^{2}\sin^{2}{\theta})\cos{\theta} \\
&-\sqrt{\frac{p(q^{2}-4m^{2})}{4m^{2}(p+q)^{3}}}[(p+q)(pr-m(p-2m))+q(p^{2}-4m^{2})]\alpha\sin^{2}{\theta}\Bigg]H_{q}^{-1}\,\rmd\phi ~,\\
\mathbf{B}  &=  -\sqrt{pq}\frac{(pq+4m^{2})r-m(p-2m)(q-2m)}{2m(p+q)\Delta_{\theta}} \alpha\sin^{2}{\theta}\,\rmd \phi ~.
\eal
\eeq
The quantities ${\bf A}$ and ${\bf B}$ can be viewed as gauge fields when performing a reduction along the $y$ and $t$-direction respectively. The associated electric and magnetic charges are\footnote{Our definition of the charges differs by a factor $8\pi$ and a minus sign for the $\mathbf{B}$-field as compared with \cite{Larsen:1999pp,Emparan:2007en}.}
\beq
\bal \label{eq:charges}
Q&= 4\pi \sqrt{\frac{q(q^2-4m^2)}{p+q}} ~,\\
P&= 4\pi \sqrt{\frac{p(p^2-4m^2)}{p+q}} ~.
\eal
\eeq
We work with positive charges, such that $p,q \geq 0$. After making the $y$-direction compact with radius $R_5$, such that $y=y+2\pi R_5$, we perform a Kaluza-Klein reduction (whose details can be found in Appendix \ref{app:KKreduction}) to obtain a rotating dyonic Kaluza-Klein black hole in four dimensions. In the Einstein frame, the action is given by
\beq
I = \frac1{16\pi G_4}\int \rmd^4x\sqrt{-g_4}\left(R_4 - \frac14 e^{-\sqrt{3}\Phi}F_{ab}F^{ab}-\frac12(\partial\Phi)^2\right) ~,
\eeq
where $\Phi$ is the canonically normalized scalar field. The equations of motion are
\beq
\bal \label{eq:4dEOM}
G_{ab} - 8\pi G_4 T_{ab} &= 0 ~, \\
\nabla_a(e^{-\sqrt{3}\Phi}F^{ab}) &= 0 ~, \\
\square\Phi + \frac{\sqrt{3}}{4}e^{-\sqrt{3}\Phi}F_{ab}F^{ab} &=0 ~,
\eal
\eeq
where the stress tensor is given by
\beq
16\pi G_4T_{ab} = e^{-\sqrt{3}\Phi}\left(F_a^{\,\,\,c}F_{bc} - \frac14g_{ab}F_{cd}F^{cd}\right) + \partial_a\Phi\partial_b\Phi - \frac12g_{ab}(\partial\Phi)^2 ~.
\eeq
By reducing the five-dimensional black hole solution we find that the dilaton is given by
\beq
\varphi = e^{-\Phi/\sqrt{3}} = \sqrt{\frac{H_q}{H_p}} ~,
\eeq
the field strength is defined as ${\bf F}=\rmd {\bf A}$ and the metric of the Kaluza-Klein black hole in the Einstein frame is
\beq
\rmd s^2 = - \frac{\Delta_\theta}{\sqrt{H_pH_q}}\left(\rmd t+{\bf B}\right)^2 + \sqrt{H_pH_q}\left(\frac{\rmd r^2}{\Delta}+\rmd\theta^2 +\frac{\Delta}{\Delta_\theta}\sin^2\theta\rmd\phi^2\right) ~.
\eeq
This metric, dilaton and field strength configuration solves the equations of motion \eqref{eq:4dEOM}. We find that the ADM mass and rotation are given by
\beq
\bal
M^{\rm KK}_4 &= \frac{p+q}{4G_4} ~, \\
J^{\rm KK}_4 &=  \frac{\sqrt{pq}(pq+4m^2)\alpha}{4G_4(p+q)m} ~.
\eal
\eeq
The horizon radii are given by the roots of $\Delta(r)$ which yields
\beq
r_\pm = m \pm \sqrt{m^2-\alpha^2} ~.
\eeq
Later, we will be interested in imposing the WGC on purely charged solutions, so with $\alpha=0$. In that case, the extremality bound is given by $m\geq 0$ which in terms of the ADM mass and charges reads
\beq
M_4^{\rm KK} \geq \frac1{16\pi G_4}(Q^{2/3}+P^{2/3})^{3/2} ~.
\eeq
By adding a $T^6$ to the four-dimensional Kaluza-Klein black hole metric these black holes can be interpreted as solutions of type IIA string theory. They then have a natural microscopic interpretation as a bound state of D0 and D6 branes \cite{Larsen:1999pp}, which give rise to the electric and magnetic charges. Denoting $N_0$ and $N_6$ as the number of D0 and D6 branes, the charges are naturally quantized in the four-dimensional theory as \cite{Emparan:2007en}
\beq
\bal \label{eq:microcharges}
Q &= \frac{16\pi G_4}{R_5}N_0~,\\
P &= 2\pi R_5 N_6~.
\eal
\eeq
We will now show how this purely charged solution is related to the rotating five-dimensional Myers-Perry black hole and the four-dimensional Kerr solution.

\subsection{Myers-Perry Black Hole}
We will now relate this rotating Kaluza-Klein solution to the Myers-Perry solution, which is the higher-dimensional generalization of the Kerr black hole \cite{Myers:1986un}. In five dimensions, its line element is given by
\beq
\bal \label{eq:MPBH}
\rmd s^2 &= -\rmd t^2 + \tilde\Sigma\left(\frac{\rho^2}{\tilde\Delta}\rmd \rho^2 + \rmd \tilde\theta^2\right) + (\rho^2+a^2)\sin^2\tilde\theta \rmd\phi^2 + (\rho^2+b^2)\cos^2\tilde\theta\rmd\tilde\psi^2 \\
 & + \frac{\mu}{\tilde\Sigma}\left(\rmd t-a\sin^2\tilde\theta\rmd\tilde\phi -b\cos^2\tilde\theta\rmd\tilde\psi\right)^2 ~,
\eal
\eeq
with
\beq
\bal
\tilde\Sigma &= \rho^2 + a^2\cos^2\tilde\theta + b^2\sin^2\tilde\theta ~, \\
\tilde\Delta &= (\rho^2+a^2)(\rho^2+b^2) - \mu \rho^2 ~.
\eal
\eeq
Here we put tildes on several quantities and coordinates that could be confused with similar expressions for the Kaluza-Klein black hole. Note that the angular coordinates obey $(\tilde\psi,\tilde\phi,\tilde\theta)=(\tilde\psi+2\pi,\tilde\phi+2\pi,\tilde\theta+\frac{\pi}{2})$.

The parameters $(\mu,a,b)$ appearing in the line element are related to the ADM mass $M_5^{\rm MP}$ and two rotations $(J_{\tilde\phi},J_{\tilde\psi})$ as follows.
\beq
\bal
M^{\rm MP}_5 &= \frac{3\pi}{8G_5}\mu ~, \\
J_{\tilde\phi} &=\frac23M_5^{\rm MP} a ~,\\
J_{\tilde\psi} &=\frac23M_5^{\rm MP} b ~.
\eal
\eeq
The horizon radii are determined by the real and positive roots of $\tilde\Delta(\rho)$ and given by
\beq
\rho_\pm = \frac{1}{\sqrt{2}} \left(\mu-a^2-b^2\pm \sqrt{\left(a^2+b^2-\mu\right)^2-4 a^2 b^2}\right)^{1/2} ~,
\eeq
Five-dimensional Myers-Perry black holes have an extremality bound given by $\mu\geq a^2+b^2+2|ab|$ which in terms of the ADM quantities is
\beq
M^{\rm MP}_5\geq \frac32\left(\frac{\pi}{4G_5}(J_{\tilde\phi}^2+J_{\tilde\psi}^2+2|J_{\tilde\phi}J_{\tilde\psi}|)\right)^{1/3} ~.
\eeq
Extremal black holes have an horizon radius given by $\rho_+ = \sqrt{|ab|}$ which, just as the area, vanishes when one of the rotations goes to zero. Thus, regular extremal black holes always have to be rotating in both rotation planes.

To obtain this solution from the Kaluza-Klein black hole we discussed previously, one can use the following identification of parameters:
\beq
\bal \label{eq:trafo1}
q &= \frac{\mu}{4p} ~,\\
\alpha &= \frac1{8p}\sqrt{\mu-(a+b)^2}(a-b) ~,\\
m &= \frac1{8p}\sqrt{\mu(\mu-(a+b)^2)} ~,
\eal
\eeq
and the following coordinate transformation:
\beq
\bal \label{eq:trafo2}
r &= \frac1{4p}\left[\rho^2-\frac12\left(\mu-a^2-b^2-\sqrt{\mu(\mu-(a+b)^2)}\right)\right] ~, \\
y &= p\psi ~.
\eal
\eeq
It is also convenient to define angular coordinates as
\beq \label{eq:trafo3}
\tilde \psi = \frac12(\psi+\phi) ~, \quad \tilde \phi = \frac12(\psi-\phi) ~, \quad \tilde\theta = \frac\theta2 ~.
\eeq
Using these relations, in the limit $p\to\infty$ the five-dimensional form of the Kaluza-Klein black hole \eqref{eq:5DKKBH} reduces to the Myers-Perry black hole \eqref{eq:MPBH}. From the periodicity $y=y+2\pi R_5$ and \eqref{eq:microcharges} we find that in this limit $\psi = \psi + 4\pi/ N_6$, such that $\tilde\psi = \tilde\psi + 2\pi/N_6$. This implies that for $N_6 > 1$ the solution is asymptotically the orbifold $\mathds{R}^4/\mathds{Z}_{N_6}$. We are interested in asymptotically flat solutions so we set $N_6=1$.

We can now also relate the different thermodynamic quantities. The mass of the Kaluza-Klein black hole diverges in the $p\to\infty$ limit. However, it can be regulated by subtracting the mass of the Kaluza-Klein monopole \cite{Emparan:2007en}. This results in 
\beq
M_5^{\rm MP} = \lim_{p\to\infty}\left(M_4^{\rm KK} - \frac{P}{16\pi G_4}\right) = \frac{3\pi}{8G_5}\mu ~.
\eeq
The two rotations of the Myers-Perry solution are related to the Kaluza-Klein charges as
\beq
J_{\tilde\psi} = \frac{PQ}{64\pi^2G_4} + J_4^{\rm KK} ~, \quad J_{\tilde\phi} = \frac{PQ}{64\pi^2G_4} - J_4^{\rm KK} ~.
\eeq
Thus, we can relate the different linear combinations of the angular momenta of the Myers-Perry black hole to the Kaluza-Klein charge and rotation respectively.
\beq
\bal \label{eq:KKMPrelation}
J_{\tilde \psi} &= \frac{PQ}{64\pi^2G_4} + J_4^{\rm KK} = \frac{N_0N_6}{2} + J_4^{\rm KK} ~,\\
J_{\tilde\phi} &= \frac{PQ}{64\pi^2G_4} - J_4^{\rm KK} = \frac{N_0N_6}{2} - J_4^{\rm KK}~.
\eal
\eeq
We note that the combination $\frac12(J_{\tilde \psi}+J_{\tilde \phi})$ is aligned along the Kaluza-Klein $y-$direction, whereas the combination $\frac12(J_{\tilde \psi}-J_{\tilde \phi})$ is aligned along $\phi$. Therefore, equal rotations $J_{\tilde \psi} = J_{\tilde \phi}$ corresponds to vanishing Kaluza-Klein rotation: $J_4^{\rm KK} = 0 $.

From the five-dimensional perspective, the Kaluza-Klein black hole can be interpreted as a black hole that is sitting at the tip of a Taub-NUT space \cite{Emparan:2006it}. This description is valid only when $N_6=1$, since the space otherwise has an orbifold singularity. As we showed, in the limit where the radius of the black hole is much smaller than the compactification radius, i.e. $R_5/r_+ \gg 1$, this black hole becomes the Myers-Perry solution.

\subsection{Kerr Black Hole} \label{eq:KerrDuality}
\begin{figure}[h]
\centering
\includegraphics[scale=.84]{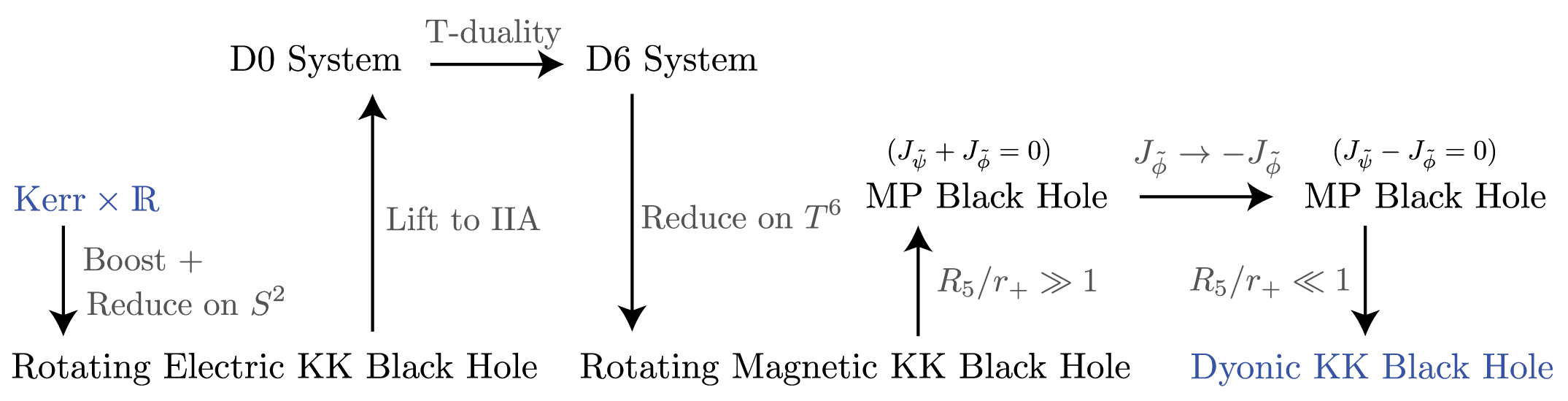}
\caption{Duality chain that maps the Kerr solution to a non-rotating dyonic Kaluza-Klein black hole. The start and endpoint of the chain are indicated in blue.}
\label{fig:KerrDuality}
\end{figure}

Finally, we relate the Kaluza-Klein and Myers-Perry solutions to the rotating Kerr black hole using the procedure of \cite{Horowitz:2007xq}. We start with the following Kerr metric.
\beq
\bal
\rmd s^2 &= \frac{\hat\Delta}{\hat\Sigma}\left(\rmd t - \hat\alpha \sin^2\hat\theta\rmd\hat\phi\right)^2 + \hat\Sigma\left(\frac{\rmd \hat r^2}{\hat \Delta}+\rmd\hat \theta^2\right) \\
&+\frac{\sin^2\hat\theta}{\hat\Sigma}\left((\hat r^2+\hat\alpha^2)\rmd\hat\phi - \hat\alpha \rmd t\right)^2 ~.
\eal
\eeq
with
\beq
\bal
\hat\Delta &= \hat r^2 - \hat r_s \hat r + \hat\alpha^2 ~, \\
\hat\Sigma &= \hat r^2 + \hat\alpha^2\cos^2\hat \theta ~.
\eal
\eeq
These quantities are related to the mass and angular momentum $(M_4^{\rm Kerr},J_4^{\rm Kerr})$ as
\beq
\hat r_s = 2G_4 M_4^{\rm Kerr} ~, \quad \hat\alpha = J_4^{\rm Kerr}/M_4^{\rm Kerr}  ~.
\eeq
The horizons are given by
\beq
\hat r_\pm = \frac12(\hat r_s \pm \sqrt{\hat r_s^2-4\hat\alpha^2}) ~,
\eeq
and the extremality bound by $\hat r_s \geq 2\hat\alpha$, which in terms of physical quantities reads
\beq
M_4^{\rm Kerr} \geq \sqrt{J_4^{\rm Kerr}/G_4} ~.
\eeq
We now take this metric, add a line to it and boost along that direction. If we now compactify on this boosted direction with radius $R_5$ we obtain a rotating Kaluza-Klein solution that carries electric charge. Taking the product with a $T^6$ we obtain a solution of type IIA string theory with D0 charge. We will consider solutions with $N_0=1$. Using T-duality along all directions of the $T^6$ (which we assume to be string scale so it does not change size) we obtain a solution with one unit of D6 charge. Reducing again on the $T^6$ we now obtain a rotating and magnetically charged Kaluza-Klein black hole in four dimensions with $N_6=1$. From our previous discussion, we know that in the limit that this black hole is small with respect to $R_5$ this becomes the Myers-Perry solution. From the relation \eqref{eq:KKMPrelation} we find that because $N_0=0$ this results in a Myers-Perry black hole with opposite rotations.
\beq
J_{\tilde \psi} = - J_{\tilde\phi} = J_4^{\rm KK} ~.
\eeq
As observed in \cite{Horowitz:2007xq}, this leads to the interesting fact that from a five-dimensional perspective, we can send $J_{\tilde\phi} \to - J_{\tilde\phi}$ such that we are now describing a Myers-Perry black hole with equal rotations. In the four-dimensional theory this exchanges Kaluza-Klein rotation for electric charge. Microscopically, we now have a D0-D6 system with $N_6=1$ and $N_0=2J_4^{\rm KK}$. We give a schematic overview of this duality chain in Figure \ref{fig:KerrDuality}.

Because the flip in angular momentum is a discrete symmetry of the Myers-Perry solution, the five-dimensional theory in the limit where the compactification radius is large cannot distinguish between a four-dimensional rotating black hole with one unit of magnetic charge and a non-rotating dyonic black hole with one unit of magnetic charge. However, their microscopic descriptions will be very different. To distinguish these two situations, the embedding of the Myers-Perry black hole in Taub-NUT needs to be known. This is an important point that will become relevant when we discuss the relation between superradiance and the WGC.

In summary, we have reviewed how the four-dimensional Kerr black hole and five-dimensional Myers-Perry black hole can both be mapped to a four-dimensional non-rotating Kaluza-Klein black hole. Because we can unambiguously impose the WGC on this non-rotating solution we can investigate how the WGC constrains the rotating solutions.

\section{Higher Derivative Corrections and the Weak Gravity Conjecture} \label{sec:HDcorrections}
We are now ready to learn how the WGC constrains extremal rotating black hole solutions. Our focus is on the mild form of the WGC which states that higher derivative corrections should increase their charge-to-mass ratio \cite{Arkani-Hamed:2006emk}. This condition ensures that even if a particle that satisfies the WGC (such as an electron) is integrated out, the resulting effective theory still has a way of allowing extremal black holes to decay by emitting smaller higher derivative corrected black holes. Initial computations of higher derivative corrections to black holes were performed by explicitly solving the higher derivative corrected equations of motion. However, by now these derivations have been greatly improved (see section 6.3 of \cite{Harlow:2022gzl} for a review and relevant references). In particular, in a canonical ensemble (fixed temperature, electric and magnetic charge) the leading correction to the extremal mass can be related to the Euclidean action as follows.\footnote{This relation is sometimes also presented as $\delta M = +T\delta I_E$. In \eqref{eq:masscorr} we made explicit the minus sign that arises from the Wick rotation to Euclidean signature.}
\beq \label{eq:masscorr}
\delta M = -\lim_{T\to 0}\left(T\int\rmd^dx\sqrt{g_d}\,\delta {\cal L}_E\right) ~.
\eeq
As usual, the time direction is periodic with $t=t+T^{-1}$, where $T$ is the temperature of the black hole and $\delta {\cal L}_E$ is the Euclidean Lagrangian containing the higher derivative corrections. We evaluate this identify on the two-derivative uncorrected solution. This expression can be obtained by noting that, in a grand canonical ensemble, the Euclidean action is equal to the Gibbs free energy: $I_E = T^{-1}G(T,\Psi_q,P)$. Here $\Psi_q$ is the electric potential and $P$ the magnetic charge. Using standard thermodynamic relations this can then be related to the mass correction of an extremal black hole in a canonical ensemble \cite{Loges:2019jzs}. The WGC is now satisfied when the correction to the mass is negative, such that the charge-to-mass ratio increases:
\beq
\text{WGC:} \quad \lim_{T\to 0}\left(T\int\rmd^dx\sqrt{g_d}\,\delta {\cal L}_E\right) \geq 0 ~.
\eeq
Using the Iyer-Wald formalism it can be shown that a necessary condition for this to happen is a violation of the dominant energy condition \cite{Aalsma:2021qga}.

Because we are considering solutions of pure gravity, the leading higher derivative corrections we have to compute involve just gravitational tensors. In five dimensions, the leading correction can be written in a basis where it is just given by the Gauss-Bonnet term. In four-dimensional pure gravity, this term is topological and the leading correction is a Riemann cubed term. Because we are considering vacuum solutions, the Gauss-Bonnet term reduces to the Riemann squared term. Thus, in total the corrections we have to study in five dimensions are
\beq \label{eq:HDcorrections}
\delta {\cal L}_E = \frac{\lambda}{L}{\cal R}^2 + \eta L{\cal R}^3 ~,
\eeq
where we used the shorthand
\beq
{\cal R}^2 = {\cal R}_{abcd}{\cal R}^{abcd} ~, \quad {\cal R}^3 = {\cal R}_{ab}^{~~cd}{\cal R}_{cd}^{~~ef}{\cal R}_{ef}^{~~ab} ~.
\eeq
$L$ is a length scale to make the Wilson coefficients ($\lambda,\eta$) dimensionless.

\subsection{Higher Derivative Corrections to Myers-Perry Black Hole}
We will first compute how the corrections \eqref{eq:HDcorrections} modify the extremality bound of the five-dimensional Myers-Perry solution. Making use of the relation \eqref{eq:masscorr} the extremal mass correction is given by
\beq
\delta M^{\rm MP}_5 = -\lim_{T\to 0}\left[T\int\rmd^5x\sqrt{g_5}\left( \frac{\lambda}{L}{\cal R}^2 + \eta L{\cal R}^3\right)\right] ~.
\eeq
Evaluating this on the Myers-Perry solution \eqref{eq:MPBH} in the extremal limit $\mu\to a^2+b^2+2|ab|$ we find
\beq
\delta M^{\rm MP}_5=  -\frac{4\pi^2\lambda}{L}\frac{a^2+b^2-6|ab|}{|ab|} - 16\pi^2\eta L \frac{\left(a^2-14 |a b|+b^2\right) \left(a^2-|a b|+b^2\right)}{7|ab|^3} ~.
\eeq
We note that for fixed Wilson coefficients, the sign of the corrections to the extremality bound depends on the ratio $|a/b|$. If one of the linear combinations of the rotations vanishes, i.e. $a\pm b=0$, we find that this expression simplifies and reduces to
\beq \label{eq:MPcorrection}
\left.\delta M^{\rm MP}_5\right|_{a\pm b=0} = \frac{16\pi^2\lambda}{L} + \frac{192\pi^2\eta L}{7a^2} ~.
\eeq
In this case, we see that the higher derivative corrections take a definite sign, depending on the sign of the Wilson coefficients. Interestingly, the limit $a=b$ is precisely the one where we can map the Myers-Perry black hole to a four-dimensional non-rotating dyonic Kaluza-Klein black hole. We will make use of this fact to constrain the higher derivative corrections to the Myers-Perry black hole.

\subsection{Higher Derivative Corrections to Kaluza-Klein Black Hole}
\begin{figure}[h]
\centering
\includegraphics[scale=1]{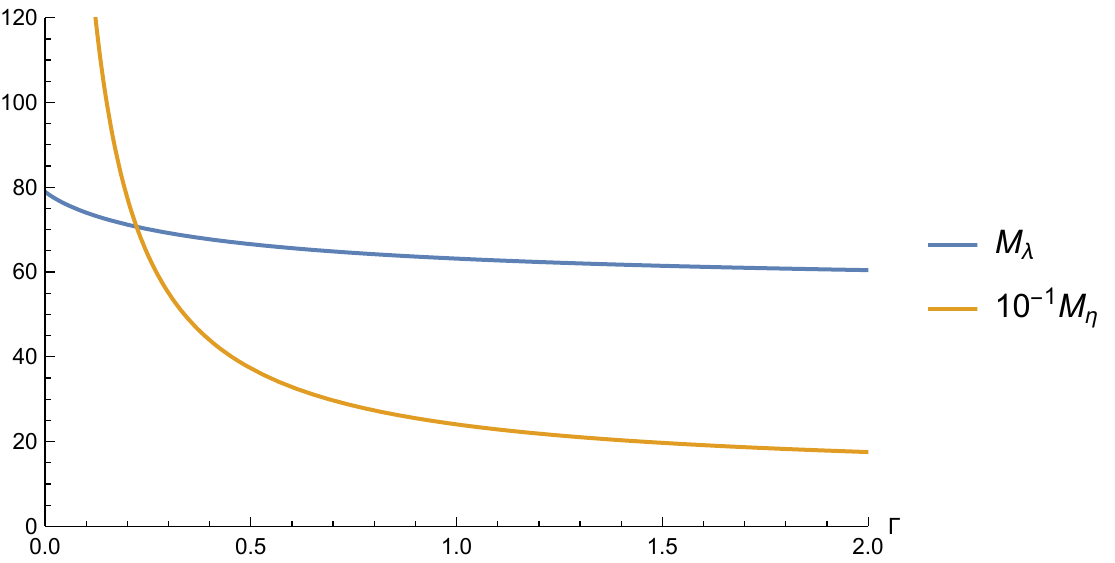}
\caption{Plot of the mass corrections \eqref{eq:MassCorrKK} as a function of the ratio $\Gamma=p/q$. These are positive function that asymptote towards a finite positive value in the limit $\Gamma\to\infty$. To make this figure we set $R_5=p=1$.}
\label{fig:MassCorrKK}
\end{figure}

We now perform the same procedure to derive the corrections to the extremality bound of the four-dimensional Kaluza-Klein solution. However, it is convenient to use the five-dimensional parent solution to compute them. The correction to the five-dimensional solution is given by
\beq
\delta M^{\rm KK}_5 = -\lim_{T\to 0}\left[T\int\rmd^5x\sqrt{g_5}\left( \frac{\lambda}{L}{\cal R}^2 + \eta L{\cal R}^3\right)\right] ~,
\eeq
where evaluate on the solution \eqref{eq:5DKKBH} in the extremal limit $m\to0$. We will now show how the four-dimensional correction can be extracted from this. By integrating over the $y$-direction we can get the four-dimensional Euclidean action.
\beq \label{eq:5Daction1}
I_E =\int\rmd^5x\sqrt{g_5}\left(\frac{\lambda}{L}{\cal R}^2 + \eta L{\cal R}^3\right) = 2\pi R_5\int\rmd^4x\sqrt{g_4}\varphi\left(\frac{\lambda}{L}{\cal R}^2 + \eta L{\cal R}^3\right) ~.
\eeq
This corresponds to the string frame action, whereas we are typically interested in the black hole solution in the Einstein frame. To remedy this, we can first perform a Weyl transformation on the five-dimensional metric. Denoting the ``old'' metric as $\tilde g_{ab}$, we transform $\tilde g_{ab} = \varphi^{-1}g_{ab}$. This results in the action
\beq \label{eq:5Daction2}
I_E = \int\rmd^5x\sqrt{g_5}\left( \varphi^{-1/2}\frac{\lambda}{L}{\cal R}^2 + \varphi^{+1/2}\eta L{\cal R}^3\right) = 2\pi R_5\int\rmd^4x\sqrt{g_4}\varphi\left( \frac{\lambda}{L}{\cal R}^2 + \varphi\eta L{\cal R}^3\right)
\eeq
The form of this Weyl transformation is chosen such that we get the four-dimensional action in the Einstein frame. We therefore find that the four-dimensional mass correction is given by
\beq
\delta M^{\rm KK}_4 = -\lim_{T\to 0}\left[T\int\rmd^5x\sqrt{g_5}\left( \varphi^{-1/2}\frac{\lambda}{L}{\cal R}^2 + \varphi^{+1/2}\eta L{\cal R}^3\right)\right] ~,
\eeq
where we evaluate on the five-dimensional metric \eqref{eq:5DKKBH} with Kaluza-Klein rotation set to zero ($\alpha = 0$). Performing the integral and taking the extremal limit $m\to 0$ we obtain the mass correction induced by the two higher derivative terms.
\beq
\delta M^{\rm KK}_4  = -\frac{\lambda}{L}{\cal M}_\lambda +\eta L {\cal M}_\eta ~.
\eeq
The form of these corrections is quite complicated and can be expressed in a relatively nice manner in terms of the ratio $\Gamma=q/p$. We then find
\beq
\bal \label{eq:MassCorrKK}
{\cal M}_\lambda  &= \frac{8\pi^2 R_5}{p}\frac{(1+\Gamma)}{(1-\Gamma)^2\sqrt{\Gamma^2-1}}\Bigg(3\pi\Gamma^2\text{sgn}(\Gamma-1)+(1-4\Gamma)\sqrt{\Gamma^2-1} \\
&-6\Gamma^2\arctan\left[\sqrt{\frac{\Gamma+1}{\Gamma-1}}\right]\Bigg) ~,\\
&\\
{\cal M}_\eta  &= \frac{16\pi^2R_5}{7p^3}\frac{(\Gamma+1)^{3/2}}{(\Gamma-1)^{9/2}\Gamma}\Bigg(105 \pi  \Gamma ^4\text{sgn}(\Gamma-1)+(6-32\Gamma+81 \Gamma ^2-160 \Gamma ^3)\sqrt{\Gamma ^2-1} \\
&-210 \Gamma ^4 \arctan\left[\sqrt{\frac{\Gamma+1}{\Gamma-1}}\right]\Bigg) ~.
\eal
\eeq
For equal charges $p=q$ these expressions simplify greatly and the mass correction is
\beq
\left.\delta M^{\rm KK}_4\right|_{p=q} = -\frac{32\pi^2R_5\lambda}{5qL} + \frac{512\pi^2R_5\eta L}{21q^3} ~.
\eeq
Importantly, the functions ${\cal M}_{(\lambda,\eta)}$ are positive along the entire domain of $\Gamma$, see Figure \ref{fig:MassCorrKK}.
In the limit $\Gamma\to\infty$ the corrections asymptote to
\beq
\bal
\lim_{\Gamma\to\infty}{\cal M}_\lambda &= \frac{4\pi^2R_5}{p}(3\pi-8) ~, \\
\lim_{\Gamma\to\infty}{\cal M}_\eta &= \frac{40\pi^2R_5}{7p^3}(21\pi-64) ~.
\eal
\eeq
In the limit where the compactification radius $R_5$ is much larger than the black hole radius, the Kaluza-Klein black hole becomes a Myers-Perry black hole. Indeed, if we use the coordinate transformation (\ref{eq:trafo1}-\ref{eq:trafo3}) on the higher derivative corrections of an extremal and non-rotating Kaluza-Klein black hole and take the limit $p\to\infty$, we find that the higher derivative corrections become
\beq
\bal
\lim_{p\to\infty}\left(\frac{\lambda}{L}{\cal R}^2 + \eta L{\cal R}^3\right) &= \frac{\lambda}{L}\frac{384 \left(3 a^8-10 a^6 \rho ^2+3 a^4 \rho ^4\right)}{\left(a^2+\rho ^2\right)^6} \\
&-\eta L\frac{12288 a^6 \left(a^6-7 a^4 \rho ^2+7 a^2 \rho ^4-\rho ^6\right)}{\left(a^2+\rho ^2\right)^9} ~.
\eal
\eeq
Then, performing the integral (with the Myers-Perry volume form) to obtain the mass correction we obtain
\beq
-\lim_{T\to 0}\left[T\int\rmd^5x\sqrt{g_5}\lim_{p\to\infty}\left(\frac{\lambda}{L}{\cal R}^2 + \eta L{\cal R}^3\right)\right] =  \frac{16\pi^2\lambda}{L} + \frac{192\pi^2\eta L}{7a^2} ~,
\eeq
which exactly matches \eqref{eq:MPcorrection}. Unlike the mass in the two-derivative theory, the correction does not need to be regulated in this limit, because the correction to the monopole mass vanishes in this limit.

\subsection{Higher Derivative Corrections to Kerr Black Hole}
Finally, we consider higher derivative corrections to the Kerr solution. The Kerr solution is more indirectly related to the Kaluza-Klein solution via the duality chain described in section \ref{eq:KerrDuality}. On the level of the action we can view it as a solution of the five-dimensional pure gravity action reducted to four dimensions with both the dilaton and electric gauge field set to zero. In four dimensions, the higher derivative corrections are then simply the Riemann squared and Riemann cubed terms. Because the Gauss-Bonnet term is topological in four dimensions, the Riemann squared term does not contribute to the correction to the extremality bound. Therefore the mass correction is solely given by
\beq
\delta M_4^{\rm Kerr} = -\lim_{T\to 0}\left[T\int\rmd^4x\sqrt{g_4} \eta L{\cal R}^3\right]  ~,
\eeq
where now ${\cal R}^3 =  R_{\mu\nu}^{~~\rho\sigma}R_{\rho\sigma}^{~~\alpha\beta}R_{\alpha\beta}^{~~\mu\nu}$. Evaluating the integral and taking the extremal limit we obtain
\beq
\delta M_4^{\rm Kerr} = \frac{8\pi\eta L}{7\hat\alpha^3} ~.
\eeq
This matches earlier results \cite{Reall:2019sah,Aalsma:2021qga}. Now that we have computed the four and six-derivative corrections to the Myers-Perry, Kaluza-Klein and Kerr black hole we can see how the WGC constrains the rotating solution. Imposing the WGC on the non-rotating, dyonic and extremal Kaluza-Klein solution fixes $\lambda\geq 0$. \footnote{Interestingly, while there is currently no known scattering positivity bound on the  Gauss-Bonnet term, the sign of $\lambda$ chosen by the WGC is consistent with that of string theory examples  \cite{Kats:2007mq,Buchel:2008vz} that violates the KSS viscosity bound \cite{Kovtun:2003wp}.} If the four-derivative term is zero (which could be the case in a particular UV-completion), the leading term is the six-derivative term. Imposing this term to decrease the mass requires $\eta \leq 0$. The constraints this places on the rotating solutions are displayed in Table \ref{tab:MassCorr}.
\begin{table}[h]
\centering
\begin{tabular}{ c | c | c }
   & $\frac{\lambda}{L}{\cal R}^2$ & $\eta L{\cal R}^3$ \\ \hline
 $\delta M_4^{\rm KK}$  & -$\frac{\lambda}{L}{\cal M}_\lambda$   & $\eta L{\cal M}_\eta$ \\
WGC:  & $\lambda\geq 0$ & $\eta\leq 0$ \\ \hline
 $\delta M_5^{\rm MP}$  & $\frac{\lambda}{L}16\pi^2$            & $\eta L\frac{192\pi^2}{7a^2}$ \\ 
Sign: & + & - \\ \hline
 $\delta M_4^{\rm Kerr}$& 0                                    & $\eta L\frac{8\pi}{7\hat \alpha^3}$ \\
Sign: & n.a. & - 
\end{tabular}
\caption{Overview of the four and six-derivative corrections to the extremal mass of the different black holes studied in this paper. The functions ${\cal M}_{(\lambda,\eta)}\geq 0$ are given in \eqref{eq:MassCorrKK}. Imposing that both higher derivative corrections decrease the mass fixes the sign of the corrections to an extremal five-dimensional Myers-Perry and four-dimensional Kerr black hole. A $+$ sign indicates an increase and a $-$ sign a decrease in mass.}
\label{tab:MassCorr}
\end{table}
The main takeaway from these results is that the sign of the corrections to the extremality bound of rotating black holes does not seem to be universal when we impose the WGC. The four-derivative correction \emph{increases} the extremal mass of the Myers-Perry solution, whereas the six-derivative correction \emph{decreases} the mass of the extremal Myers-Perry and Kerr solutions. We give an interpretation of this non-universal behavior in terms of the instability of these black holes in the next section.

\section{Superradiant Instabilities} \label{sec:superradiance}
From the results in Table \ref{tab:MassCorr} it is clear that the higher derivative terms we considered do not correct the extremality bound of extremal Kerr and five-dimensional Myers-Perry black holes with a universal sign. In that sense, there does not seem to be a sharp statement, like the charged WGC, that requires corrections to extremal rotating black holes to increase their angular momentum-to-mass ratio.\footnote{As mentioned before, BTZ black holes are an exception. When corrections to the extremality bound can be viewed as relevant perturbations in the CFT the holographic c-theorem can be used to argue for a spinning WGC \cite{Aalsma:2020duv}.}

At first sight, one might not be surprised by this observation. As discussed in the introduction, one of the motivations for the charged WGC, extremal black hole decay, does not hold for rotating black holes due to superradiance. A necessary condition for superradiance to occur in rotating black holes is the presence of an ergosphere where the timelike Killing vector flips sign. In the extremal limit, the radius of the ergosphere is
\beq
\bal
\text{Kerr:}\qquad r_e  &= \hat\alpha(1+\sin\theta) \geq r_+ ~,\\
\text{Myers-Perry:}\qquad \rho_e  &= \sqrt{a^2\sin^2\theta + b^2\cos^2\theta + 2|ab| } \geq \rho_+  ~.
\eal
\eeq
Because the radius of the ergosphere lies outside of the black hole horizon, this allows the presence of negative energy states that can be used to extract energy and angular momentum by scattering a particle off a black hole. This also implies that, quantum mechanically, these modes can be spontaneously created. From this perspective, no WGC-like constraint is expected for rotating black holes. On the other hand, the rotating black holes we study can be mapped to purely charged solutions. Charged extremal black holes do not have an ergoregion and are typically stable if we do not impose the WGC. This therefore raises the question how the superradiant instability of the rotating solutions arises in the charged solutions. We will now show in detail that, in order for the superradiant instability to exist, the charged solutions need to satisfy the WGC. A similar observation about the spectrum of charged states was made in \cite{Emparan:2007en}, but here we demonstrate the relationship with the WGC.

\subsection{Rotational Superradiance}
The condition for superradiance to occur can be derived in an elegant manner from black hole thermodynamics \cite{Brito:2015oca}. The first law for rotating black holes is 
\beq
\rmd M = T\rmd S + \Omega_i \rmd J^i ~.
\eeq
Here $\Omega_i$ is the angular potential and the index $i$ runs over different possible angular momenta. The potentials are given by
\beq
\Omega_\phi = \frac{\hat r_s-\sqrt{\hat r_s^2-4 \hat\alpha ^2}}{2 \hat\alpha \hat r_s} ~, \quad \Omega_{\tilde\phi} = \frac a{\rho_+^2+a^2} ~, \quad \Omega_{\tilde\psi}=\frac b{\rho_+^2+b^2} ~.
\eeq
Let's say that we perform a scattering experiment and want to extract energy $\omega$ and angular momentum $j^i$ from the black hole such that $\rmd M/\rmd J^i  = \omega/j^i$. Using the first law we can then write
\beq
\rmd M = T\rmd S \frac{\omega}{\omega - j^i\Omega_i} ~.
\eeq
Using the second law ($\rmd S\geq 0$) and imposing we extract energy ($\rmd M \leq 0$) we then find that the condition for superradiance to occur is
\beq \label{eq:SRcond}
\omega \leq j^i\Omega_i ~.
\eeq
The extremal limit should be taken after this condition has been derived and not directly in the first law. Evaluating the angular potential in the extremal limit for the solutions of interest we find
\beq
\bal 
\text{Kerr:}\qquad \Omega_\phi  &= \frac1{2\hat \alpha} ~,\\
\text{Myers-Perry:}\qquad \Omega_{(\tilde\psi,\tilde\phi)} & = \frac1{a+b} ~.
\eal
\eeq
We can now compare the superradiance condition \eqref{eq:SRcond} against a possible spinning WGC bound. These conditions can be obtained from the extremality bound and are given by
\beq
\bal \label{eq:spinWGC}
\text{WGC for Kerr:}&\qquad \omega\sqrt{\frac{G_4}{j^\phi}}  \leq 1 ~,\\
\text{WGC for Myers-Perry:}&\qquad \frac23 \left(\frac{4G_5}{\pi}\right)^{1/3}\frac{\omega}{\left((j^{\tilde\psi})^2+(j^{\tilde\phi})^2 +2|j^{\tilde\psi}j^{\tilde\phi}|\right)^{1/3}} \leq 1 ~.
\eal
\eeq
Comparing this spinning WGC bound with the condition for superradiance, we see that it is possible to obey \eqref{eq:SRcond} while not satisfying \eqref{eq:spinWGC}. From the perspective of the rotating black holes, there  is thus no clear relationship between the onset of superradiance and the extremality bound. However, as we will now show this behaviour is different for charged black holes.

\subsection{Charged Superradiance}
Charged black holes have no ergosphere, so in a strict sense they do not superradiate. However, they can lose mass and charge when a similar condition as \eqref{eq:SRcond} is satisfied and we will refer to this as charged superradiance. We will see that in this case there is a clear relation between the WGC and the condition for superradiance to occur. This was observed earlier in \cite{Aalsma:2018qwy}. The first law for dyonic black holes is given by
\beq
\rmd M = T\rmd S + \Psi_q \rmd Q + \Psi_p \rmd P ~. 
\eeq
Here $\Psi_{q,p}$ are the electric and magnetic potentials and $(Q,P)$ the electric and magnetic charges. In the extremal limit the potentials are
\beq
16\pi G_4\Psi_q = \left.A_t\right|^{r=\infty}_{r=r_+} = \sqrt{\frac{p+q}{q}} ~, \quad 16\pi G_4\Psi_p = \sqrt{\frac{p+q}{p}} ~,
\eeq
where we obtained the magnetic potential from the electric one by sending $q\leftrightarrow p$. Let us consider extracting energy $\omega$ and electric and magnetic charge $(k_q,k_p)\geq 0$. This results in the following change in black hole parameters
\beq
\frac{\rmd M}{\rmd Q} = \frac{16\pi G_4\omega}{k_q} ~, \quad \frac{\rmd M}{\rmd P} = \frac{16\pi G_4\omega}{k_p} ~.
\eeq
We normalized the charges $k_{q,p}$ in such a way that $k_q$ corresponds to the integer quantized electric charge in units of $R_5$. Following the same steps as before (imposing the second law and requiring $\rmd M \leq 0$) we can derive the following condition for charged superradiance:
\beq
16\pi G_4\omega \leq k_q\Psi_q + k_p\Psi_p ~.
\eeq
We find that the conditions for superradiance and the WGC are given by
\beq
\bal
\text{Superradiance:} &\quad \frac{16\pi G_4\omega}{k_q\sqrt{1+(P/Q)^{2/3}}+k_p\sqrt{1+(Q/P)^{2/3}}} \leq 1 ~, \\
\text{WGC:} &\quad \frac{16\pi G_4\omega}{(k_q^{2/3}+k_p^{2/3})^{3/2}} \leq 1 ~.
\eal
\eeq
It is straightforward to check that when the superradiance condition is satisfied, this implies the WGC condition (but not the other way around).
\beq
\frac{16\pi G_4\omega}{k_q\sqrt{1+(P/Q)^{2/3}}+k_p\sqrt{1+(Q/P)^{2/3}}} \leq  \frac{16\pi G_4\omega}{(k_q^{2/3}+k_p^{2/3})^{3/2}} ~.
\eeq
The superradiance mass-to-charge ratio is maximized when $k_p/k_q = P/Q$, where it equals the WGC condition. We note that if we were to consider a solution that is purely electric or magnetic, the superradiance and WGC condition are always equivalent.

\subsection{Superradiance vs. Weak Gravity Conjecture}
Although for the rotating black holes we considered in this paper there is no obvious relation between the condition for superradiant emission and the extremality bound, the situation is qualitatively different for charged solutions. Whenever the extremal Kaluza-Klein black hole has a charged superradiant instability, the superradiant states also satisfy the WGC. This gives an interesting perspective on the superradiant instability for the extremal rotating solutions we considered in this paper. As discussed previously, from the perspective of the five-dimensional Myers-Perry black hole with equal rotations, it is not clear if in four-dimensions this is described by a rotating black hole with one unit of magnetic charge or a non-rotating dyonic black hole. The former solution naturally has a superradiant instability without a strong constraint on the particle spectrum, whereas the second solution can only have a superradiant instability when the WGC is satsified. Thus, whenever the five-dimensional Myers-Perry black hole is described in four-dimensions by a dyonic solution, its superradiant instability implies the WGC in four dimensions.

With this knowledge, it is perhaps not unexpected that the signs of the higher derivative corrections in Table \ref{tab:MassCorr} do not conform to a spinning WGC. The rotating extremal black holes are already unstable without the need of having superextremal rotating particles. When we can map these rotating solutions to pure charge black holes, this instability manifests itself as the ordinary charged WGC. What is interesting, is that the WGC can still be used to constrain the Wilson coefficients of higher derivative operators correcting the extremality bound.

\section{Conclusions} \label{sec:conclusion}
In this paper, we computed the leading higher derivative corrections to extremal rotating black holes. We focused on the four-dimensional Kerr and five-dimensional Myers-Perry black hole, both of which can be mapped to a four-dimensional non-rotating dyonic Kaluza-Klein black hole. By imposing the WGC on this purely charged solution we fixed the sign of the Wilson coefficients and, with that, the sign of the corrections to the extremality bound of the rotating solutions. This way, the WGC leads to new bounds that extremal rotating black holes must satisfy to belong to the landscape. 

The sign of the corrections, however, does not seem to be universal. While the extremal mass of the Kerr solution decreases, the mass of the Myers-Perry solution increases. We gave an interpretation of this non-universality in terms of the instability of rotating black holes: unlike charged extremal solutions, rotating black holes have a superradiant instability that allows them to shed their angular momentum without a constraint on the spectrum that is directly related to the extremality bound. Nonetheless, the mapping of rotating to charged black holes shows that this superradiant instability can only be satisfied when the charged WGC is satisfied.

It would therefore be interesting to understand if there are extremal rotating black holes that do not exhibit superradiance. If the heuristic argument that extremal black holes should be able to decay is correct, we expect a stronger constraint on the spectrum of spinning states when considering such black holes. Indeed, there are extremal rotating (and charged) solutions such as the BMPV black hole that are stable because they saturate a BPS bound \cite{Breckenridge:1996is}. However, precisely because of their BPS-ness, corrections to the extremality bound vanish \cite{Liu:2022sew} saturating a possible WGC-like constraint. One is therefore led to the question if there exist non-BPS solutions that do not exhibit superradiance.

Furthermore, recently there has been a sharpened understanding of the constraints that Wilson coefficients of higher derivative corrections to pure gravity, like the ones considered in this paper, need to satisfy (see e.g. \cite{deRham:2021bll,Chen:2021bvg,Guerrieri:2021ivu,Caron-Huot:2021rmr,Caron-Huot:2022ugt}). Here, we assumed the WGC to obtain constraints on rotating black holes but it would be interesting to see if such techniques can provide an independent way of determining the sign of the Wilson coefficients of gravitational operators. Turning the argument around, if certain swampland constraints are established, we can use them in combination with duality (as we have done in this paper) to obtain new positivity bounds that would otherwise be difficult to prove directly with amplitude techniques. Hopefully, these diverse efforts will lead to a sharper understanding of the physical principles that separate the landscape from the swampland.

\section*{Acknowledgments}
This work is supported in part by the DOE under Grant No. DE-SC0017647. We gratefully acknowledge the hospitality of the Center of Mathematical Sciences and Applications and the Physics Department at Harvard University where parts of this work were completed. GS would additionally like to thank Toshifumi Noumi and the participants of the ``Possible and Impossible in Effective Field Theory: From the S-Matrix to the Swampland" Workshop at the Institute for Advanced Study in May 2022 for discussions.

\appendix

\section{Details on Kaluza-Klein Reduction} \label{app:KKreduction}
We want to take the action of pure five-dimensional gravity perturbed by higher derivative corrections up to six derivatives and perform a Kaluza-Klein reduction. The relevant action is given by
\beq
I = \int\rmd^5x\sqrt{-g_5}\left(\frac{R_5}{16\pi G_5} + \frac{\lambda}{L}{\cal R}^2 + \eta L{\cal R}^3\right) ~.
\eeq
We use the following short-hand notation for the higher derivative corrections.
\beq
{\cal R}^2 = {\cal R}_{abcd}{\cal R}^{abcd} ~, \quad {\cal R}^3 = {\cal R}_{ab}^{~~cd}{\cal R}_{cd}^{~~ef}{\cal R}_{ef}^{~~ab} ~.
\eeq
Here $R_5$ is the five-dimensional Ricci scalar ${\cal R}_{abcd}$ is the five-dimensional Riemann tensor.
To perform the Kaluza-Klein reduction, we assume the standard ansatz
\beq
\rmd s^2 = g_{\mu\nu}\rmd x^\mu\rmd x^\nu + \varphi^2(A_\mu\rmd x^\mu + \rmd y)^2 ~,
\eeq
where $g_{\mu\nu}$ is the four-dimensional metric and $y=y+2\pi R_5$. Note that roman indices run as $a=(0,\dots,4)$ and greek indices as $\mu=(0,\dots,3)$. In order to derive the various expressions for the geometric tensors in the four-dimensional theory, it is useful to first go to a flat basis using the vielbeins $e^a_{\,\,\,\hat a}$ such that
\beq
g_{ab} = e^{\hat a}_{\,\,\,a}e^{\hat b}_{\,\,\,b}\eta_{\hat a\hat b} ~,
\eeq
where $\eta_{\hat a\hat b}$ is the Minkowski metric.

In this flat basis, the different components of the five-dimensional Riemann tensor are given by \cite{Wehus:2002se}
\beq
\bal \label{eq:RiemmExpress}
\hat {\cal R}_{\mu\nu\rho\sigma}  &= \hat R^{(4)}_{\mu\nu\rho\sigma} - \frac14\varphi^2\left(2\hat F_{\mu\nu}\hat F_{\rho\sigma}+ \hat F_{\mu\rho}\hat F_{\nu\sigma}-\hat F_{\mu\sigma}\hat F_{\nu\rho}\right) ~, \\
\hat {\cal R}_{4\mu\nu\rho} &= \frac12\varphi\hat \nabla_\mu \hat F_{\nu\rho} + \frac12\left(2(\hat \partial_\mu\varphi)\hat F_{\nu\rho} + (\hat \partial_\nu\varphi)\hat F_{\mu\rho}-(\hat \partial_\rho\varphi)\hat F_{\mu\nu}\right) ~,\\
\hat {\cal R}_{\mu4\nu4} &= -\varphi^{-1}\hat \nabla_\mu(\hat \partial_\nu\varphi) -\frac14\varphi^2\hat F_{\mu\rho}\hat F^\rho_{\,\,\,\nu} ~.
\eal
\eeq
Here we used the hat to indicate these expressions are valid in the flat basis. From these expressions, the components of the five-dimensional Ricci tensor can also be constructed.
\beq
\bal
\hat {\cal R}_{\mu\nu} &= \hat R^{(4)}_{\mu\nu}-\frac12\varphi^2 \hat F_\mu^{\,\,\,\rho}\hat F_{\nu\rho} - \frac12\varphi^{-2}\hat \nabla_\mu(\hat\partial_\nu\varphi^2) + \varphi^{-2}(\hat \partial_\mu\varphi)(\hat \partial_\nu\varphi) ~, \\
\hat {\cal R}_{\mu4} &= \frac12\varphi^2\hat \nabla_\nu\hat F_\mu^{\,\,\,\nu} + \frac32\hat F_\mu^{\,\,\,\nu}\hat \partial_\nu\varphi ~, \\
\hat {\cal R}_{44} &= \frac14\varphi^2 \hat F_{\mu\nu}\hat F^{\mu\nu} - \frac12 \varphi^{-2}\hat \square\varphi + \varphi^{-1}(\hat \partial_\mu\varphi)(\hat \partial^\mu\varphi) ~.
\eal
\eeq 
We can transform these expressions back to the non-flat basis using the vielbeins. Explicitly, in the directions we compactify over they are
\beq
e^{\hat \mu}_{\,\,\,4} = 0 ~, \quad e^{\hat 4}_{\,\,\,\mu} = \varphi A_\mu ~, \quad e^{\hat 4}_{\,\,\,4} = \varphi ~.
\eeq
Then, by appropriately contracting the expressions \eqref{eq:RiemmExpress} we can obtain four-dimensional expressions for the higher derivative corrections ${\cal R}^2$ and ${\cal R}^3$. For our purposes, it will be enough to leave them in their five-dimensional form. The four-dimensional expression for the Ricci scalar is
\beq
R_5 = R_4 - \frac14\varphi^2F_{\mu\nu}F^{\mu\nu}  -\varphi^{-2}\square\varphi^2 + 2\varphi^{-2}(\partial_\mu\varphi)(\partial^\mu\varphi) ~.
\eeq
Integrating over the $y$ direction, the four-dimensional action becomes
\beq
I = \frac1{16\pi G_4}\int\rmd^4x\sqrt{-g_4}\varphi\left(R_4 - \frac14\varphi^2F_{\mu\nu}F^{\mu\nu} \right) + 2\pi R\int \rmd ^4x\sqrt{-g_4} \varphi \left(\frac{\lambda}{L}{\cal R}^2 + \eta L {\cal R}^3\right)
\eeq
The four-dimensional Newton constant is $G_4 = G_5 /(2\pi R_5)$. This is the action in the string frame. Performing a Weyl transformation we can bring it to the Einstein frame. Under a transformation of the form $\tilde g_{ab} = e^{2\omega}g_{ab}$ the relevant quantities transform as
\beq
\bal
\sqrt{-\tilde g_d} &= e^{d\omega}\sqrt{-g_d} ~, \\
\tilde R_d &= e^{-2\omega}\Big(R_d - 2(d-1)\square\omega - (d-2)(d-1)(\partial_a\omega)(\partial^a\omega)\Big) ~,\\
\tilde {\cal W}^a_{\,\,\,bcd} &= {\cal W}^a_{\,\,\,bcd} ~.
\eal
\eeq
Here ${\cal W}^a_{\,\,\,bcd}$ is the Weyl tensor, which is invariant under Weyl transformations. Because we are interested in vacuum solution in five dimensions, the five-dimensional Riemann tensor is equal to the five-dimensional Weyl tensor. This implies that the higher derivative terms transform covariantly under Weyl transformations as
\beq
\bal
\tilde {\cal R}^2 &= e^{-4\omega}{\cal R}^2 ~,\\
\tilde {\cal R}^3 &= e^{-6\omega}{\cal R}^3 ~.
\eal
\eeq
Identifying $\tilde g_{\mu\nu}$ as the string frame metric and taking $e^{2\omega}\varphi=1$ puts the action in the Einstein frame. Finally, introducing a canonically normalized scalar field $\omega =\Phi/\sqrt{12}$ we obtain the final form of the action
\beq
\bal
I = \frac1{16\pi G_4}&\int\rmd ^4x\sqrt{-g_4}\left(R_4 - \frac12(\partial_\mu\Phi)(\partial^\mu\Phi)-\frac14 e^{-\sqrt{3}\Phi}F_{\mu\nu}F^{\mu\nu}\right) \\
+ 2\pi R_5&\int\rmd ^4x\sqrt{-g_4}\left(\frac{\lambda}{L}e^{-\Phi/\sqrt{3}}{\cal R}^2+\eta L e^{-2\Phi/\sqrt{3}}{\cal R}^3 \right) ~.
\eal
\eeq

\bibliographystyle{JHEP}
\bibliography{refs}

\end{document}